\documentclass[12pt]{article}
\usepackage[T2A]{fontenc}
\usepackage[tbtags]{amsmath}
\usepackage{amsfonts,amssymb,mathrsfs,amscd}



\usepackage{amsmath}
\usepackage{amsfonts}
\usepackage{amssymb}
\usepackage{amscd}
\usepackage{amsthm}
\usepackage{bbm}
\usepackage{color}
\usepackage[linktocpage=true,plainpages=false,pdfpagelabels=false]{hyperref}

\usepackage[matrix,arrow,curve]{xy}

\begin{document}
\title{
Lax Pairs for Linear Hamiltonian Systems\thanks{The work of the first author   is partially supported by Laboratory of Mirror Symmetry NRU HSE, RF Government grant, ag.~\textnumero 14.641.31.0001. The work of the second author was supported by RFBR grants 16-01-00378-a and 16-51-55012 China-a.}
}

\author{
D.~V.~Osipov, A.~B.~Zheglov
}

\date{}
\maketitle

\begin{abstract}

In the paper, Lax pairs for linear Hamiltonian systems of differential equations are constructed. In particular, Gr\"obner bases are used for the computations. It is proved that the maps which appear in the construction of Lax pairs are Poisson. Various properties of first integrals of the system which are obtained from  the Lax pairs are investigated.


\end{abstract}

\section{Introduction}

The aim of this paper is to calculate  Lax pairs for a linear system of differential equations
$$
\Gamma \dot{x}= - Px  \, \mbox{,} \eqno (1),
$$
where the column vector
$x \in  {\mathbb{R}}^{2n}$,
matrices
$\Gamma \in {\rm M}_{2n}({\mathbb{R}})$,
$P \in {\rm M}_{2n}({\mathbb{R}})$.
Here the matrix $P$ is assumed to be symmetric, and the matrix $\Gamma$ is assumed to be skew-symmetric, that is,
$$P^{\rm T}=P \quad \mbox{and} \quad {\Gamma}^{\rm T}= - {\Gamma} \mbox{.}$$
We will also assume that
$\det{\Gamma} \ne 0$
(what explains the parity of the dimension of the vector space) and $\det{P}  \ne 0$.

On the other hand, if we consider an arbitrary system of differential equations
$$
\dot{x} = Ax  \, \mbox{,}\eqno (2)
$$
admitting a quadratic first integral $\frac{1}{2} x^{{\rm T}} P x$,  where the column vector $x \in {\mathbb{R}}^r$, matrices
$A \in {\rm M}_{2n}({\mathbb{R}})$, $P \in {\rm M}_{2n} ({\mathbb{R}})$, $P^{{\rm T}}=P$ and $\det A \ne 0$, $\det P \ne 0$,
then
$\Gamma= - PA^{-1}$ is a non-degenerate skew-symmetric matrix (see works [1,\,2]),
and  system~(2) is immediately reduced to the form~(1), therefore $r =2n$.

We note that system~(1) is Hamiltonian, see  section~2 below.

By {\em a Lax pair of dimension} $ k $ for system~(1) we mean a smooth map
$$
{\rm LP}  \; :  \; \mathbb{R}^{2n}  \longrightarrow {\rm M}_k(\mathbb{C}) \times {\rm M}_k(\mathbb{C})  \simeq \mathbb{R}^{4k^2}  \, \mbox{,}\eqno (3)
$$
which translates the solutions of  system~(1) into the solutions of the following matrix differential equation:
$$
\dot{L} = [B, L]  \, \mbox{,}\eqno (4)
$$
where $B \times L \in {\rm M}_k(\mathbb{C}) \times {\rm M}_k(\mathbb{C}) $.
In other words, if $\tau$ is a vector field on ${\mathbb R}^{2n}$ generated by system~(1), then for any $x \in {\mathbb R}^{2n}$ we have that the $ L $-component of the tangent vector
${\rm d LP}_x (\tau(x))$ at the point ${\rm LP}(x)$ is defined by the system~(4).

From here, as it is known, it follows that for the function
 $$f_l = {{\mathop {\rm Tr}}} (L^l)  \; : \; {\rm M}_k(\mathbb{C})  \longrightarrow \mathbb{C}  \, \mbox{,} \qquad \mbox{where} \quad l \in {\mathbb N} \, \mbox{,}$$
  holds $\dot{f}_l=0$, that is, the function $f_l$ is constant on the solutions of the system~(4). Therefore, the preimage of any functional combination of the real and imaginary parts of the function $f_l$ with respect to the map  ${\rm pr}_2 \circ {\rm LP}$ is the first integral of the system~(1). The main problem in this approach is to verify the nontriviality of the integrals, as well as the question of the completeness of the system of first integrals obtained using different Lax pairs.

We note that the study of Lax pairs for linear Hamiltonian systems is the first step to the non-linear case, since any non-linear system of differential equations gives a linear system in the first approximation. For example, if the system is Hamiltonian, then one can decompose the Hamiltonian of the system into a Taylor series and leave for further consideration only terms with order of degree at most two.

 For $ n = 1 $, we  investigate all possible Lax pair of dimension $ 2 $ for the system~(1) in section~3. For the final answer in this case, we also use computer calculations using the Gr\"obner bases. The use of Gr\"obner bases for finding Lax pairs, as far as we know, has never been used before this work. They were also used to get Lax pairs in the general case (see below). For a brief overview of the technology associated with the Gr\"obner bases, see, for example, in the brochure~[3] or in the book~[4].

 For an arbitrary $n \ge 1$, as well as for a pair: an eigenvalue and an eigenvector for this eigenvalue of the matrix $(P \Gamma^{-1})^2$,
we build a Lax pair of dimension $ 2 $ in section~4. We show that for different eigenvalues the first integrals of the system~(1), constructed from the function~${\mathop {\rm Tr}} (L^2)$, are in involution on the space ${\mathbb{R}}^{2n}$ with respect to the Poisson bracket given by the skew-symmetric matrix $-\Gamma$. These first integrals are quadratic functions on the space ${\mathbb{R}}^{2n}$, and in the case of a simple spectrum of the matrix $P \Gamma^{-1}$ forms a complete set of $n$ functionally independent first integrals of Hamiltonian system~(1) (for complex eigenvalues and eigenvectors, we must take the real and imaginary parts of the corresponding functions, as we noted above). Thus, we obtain a new simple proof of the result of J.~Williamson, see [5], that the system~(1) admits $ n $ first quadratic integrals that are pairwise in involution and functionally independent, see Remark 4.3. (A detailed discussion of this theme, with additional links, is provided at the end of~[2, \S\,2].)

 In Section~5, we show that the map arising in the construction of Lax pairs is a morphism of Poisson manifolds with respect to the natural Poisson brackets.

We are grateful to A.~N.~Parshin for posing the problem and numerous discussions.
The starting point of our research was the report by V.~V.~Kozlov "The Symplectic Geometry of Linear Hamiltonian Systems and the Solution of Algebraic Equations" at the joint seminar of the departments of Algebra and  Algebraic geometry at Steklov Mathematical Institute in September 2017.

\section{Preliminaries}

Recall the following well-known facts about symplectic structures and Poisson brackets.

Let a non-degenerate skew-symmetric matrix $W \in {\rm M}_s({\mathbb{R}})$, where $s$ is an even number, defines a symplectic structure on the space ${\mathbb{R}}^s$ with the help of {$2$-form}
$$\Omega(x,y) = x^{{\rm T}} W y \quad \mbox{,}  \quad x, y \in {\mathbb{R}}^s \, \mbox{.}$$
The vector field $ v $ on ${\mathbb{R}}^s$ is called Hamiltonian with Hamiltonian $H$, which is a smooth function on ${\mathbb{R}}^s$, if
$$\Omega(v,w)= -dH(w)  \, \mbox{,}$$
where $w$ is an arbitrary vector field on  ${\mathbb{R}}^s$.
We will denote such a vector field as ${v = \mathop{\rm sgrad} H}$. For two smooth functions $ f $ and $ g $ on the space ${\mathbb{R}}^s$ their Poisson bracket $\{f,g  \}$ is determined by the following equations:
$$
\{f,g\}= - \Omega(\mathop{\rm sgrad} f, \, \mathop{\rm sgrad} g )= -dg (\mathop{\rm sgrad} f)= df(\mathop{\rm sgrad} g)  \mbox{.}
$$
Then, in coordinates,
$$
\mathop{\rm sgrad} g = W^{-1} \cdot \mathop{\rm grad} g \quad \mbox{and}  \quad \{f,g \} = (\mathop{\rm grad} f)^{{\rm T}} \cdot  W^{-1} \cdot \mathop{\rm grad} g \, \mbox{,}\eqno (5)
$$
where the column of functions, the gradient, is $\mathop{\rm grad} f = (\frac{\partial f}{\partial x_1}, \ldots, \frac{\partial f}{\partial x_s})^{{\rm T}}$,
and $x_1, \ldots, x_s$ are standard coordinates in ${\mathbb{R}}^s$.

Returning to the system ~ (1) (in this case $ s = 2n $), one can see by simple calculations that if the matrix $W= -\Gamma $, then the vector field
 $- \Gamma^{-1} P x$ is Hamiltonian with Hamiltonian $H = \frac{1}{2}x^{{\rm T}} P x$, $x \in {\mathbb{R}}^{2n}$,
see also~[2, \S\,1].
Therefore, the system~(1) is Hamiltonian with respect to the symplectic structure described above on ${\mathbb{R}}^{2n}$, specified by the skew-symmetric matrix $-\Gamma$.

Further we will use the following statements. We assume that the matrices $ \ Gamma $ and $ P $ satisfy the conditions, as at the beginning of the Introduction.
\smallskip

{\bf \noindent {Proposition 2.1.}}  {\it
The following is true.
\begin{enumerate}
\item  If $\lambda  \in {\mathbb{C}}$ is an eigenvalue of the matrix $P \Gamma^{-1}$, then $\overline{\lambda}$, $-\lambda$, $- \overline{\lambda}$ are also eigenvalues of the matrix $P \Gamma^{-1}$.
\item  Let $v_1$ and $v_2$ be (complex) eigenvalues of the matrix  $(P \Gamma^{-1})^2$  with corresponding eigenvalues $\lambda_1$ and $\lambda_2$  from ${\mathbb{C}}$. If $\lambda_1 \ne \lambda_2$, then
    $$ v_1^{{\rm T}} \Gamma^{-1} v_2  =0 \, \mbox{.}
    $$
\item  Let the spectrum of the matrix $P \Gamma^{-1}$  be simple. If $ v $ is an eigenvector of the matrix $(P \Gamma^{-1})^2$, which is not an eigenvector of the matrix $P \Gamma^{-1}$,
    then
    $$
    v^{{\rm T}} \Gamma^{-1} (P \Gamma^{-1} v)  \ne 0 \, \mbox{.}
    $$
\end{enumerate}
}
\begin{proof}

1. This property follows from the fact that the matrix $P \Gamma^{-1}$ is real, and from the equalities
$$
0 = \det (P \Gamma^{-1}  - \lambda E)= \det (P - \lambda \Gamma) = \det(P - \lambda \Gamma)^{{\rm T}} = \det (P + \lambda \Gamma)= \det(P \Gamma^{-1} + \lambda E) \, \mbox{,}
$$
where $E$ is the identity matrix.

2. This property follows from the chain of equalities:
\begin{multline*}
 v_1^{{\rm T}} \Gamma^{-1} v_2 = (\lambda_1^{-1} (P \Gamma^{-1})^2 v_1)^{{\rm T}} \Gamma^{-1} v_2 =
 \lambda_1^{-1} v_1^{{\rm T}} \Gamma^{-1} P \Gamma^{-1} P \Gamma^{-1} v_2 = \\ = \lambda_1^{-1} v_1^{{\rm T}} \Gamma^{-1} (P \Gamma^{-1})^2 v_2
 =
 \lambda_1^{-1}  \lambda_2  v_1^{{\rm T}} \Gamma^{-1} v_2  \mbox{.}
\end{multline*}

3. Since the spectrum of the matrix $P \Gamma^{-1}$ is simple, then from item~1 of this proposition it follows that the matrix $(P \Gamma^{-1})^2$
is diagonalized, with a two-dimensional subspace of eigenvectors corresponding to each eigenvalue. Let $ D $ be a matrix composed of columns that are linearly independent (over $\bf {\mathbb{C}}$) eigenvectors of the matrix $(P \Gamma^{-1})^2$, so that the columns corresponding to one eigenvalue are adjacent. Then the matrix $D^{{\rm T}} \Gamma^{-1} D$ has the number $v_l^{{\rm T}} \Gamma^{-1} v_j$ on the $ (l, j) $ place, where $v_k$ is the $k$-th column of $D$. Therefore, from item~2 of this proposition, we obtain that $G=D^{{\rm T}} \Gamma^{-1} D$ is a block-diagonal matrix with $ 2 \times 2 $ blocks on the diagonal.  Note that $P \Gamma^{-1} v$ is also an eigenvector of the matrix  $(P \Gamma^{-1})^2$  with the same eigenvalue as the vector $v$,
but the vector $P \Gamma^{-1} v$ is not proportional to the vector $v$. Therefore, if  $ v^{{\rm T}} \Gamma^{-1} (P \Gamma^{-1} v) =0$, then one of the $ 2 \times 2 $ blocks of the matrix  $ G $  would be equal to zero.  It would follow from this that $\det G =0$. Therefore,  $\det \Gamma^{-1}=0$, a contradiction.
\end{proof}

\section{Lax Pairs for $ n = 1 $}
\subsection{Using the square root of the matrix}

In this section, we present the Lax pair of dimension $ 2 $ for the system~(1) in the case $ n = 1 $, so that the function ${\mathop {\rm Tr}} (L^2)$ will be the first integral of this system coinciding with $ 4H $, where the Hamiltonian is  $H = \frac{1}{2} x^{{\rm T}} P x$. Later, in section~4 we give the general formulas for Lax pairs of dimension $ 2 $ for system~(1), but these formulas will differ from this one in the case of $ n = 1 $ and the Hamiltonian $ H $ (for arbitrary $ n $) will be obtained only as a linear combination of the first integrals connected with Lax pairs.

Since $P^{{\rm T}} =P$, there exists a symmetric matrix $T \in {\rm M}_{2}({\mathbb{C}})$, such that  $T^2=P$. (Such a matrix always exists, since by conjugation by orthogonal matrices from ${\rm O}(2,{\mathbb{R}})$ the matrix $P$ can always be reduced to a diagonal matrix, from which we take the root in the form of a diagonal matrix and then apply the reverse pairing.) Note that $\det T \ne 0$.

We will use the following identity
$$
\Gamma \cdot T = \det T \cdot T^{-1} \cdot \Gamma  \, \mbox{,}\eqno (6)
$$
which is immediately verified by direct calculations for explicit matrices
$$
\Gamma = \left(\begin{array}{cc}
0&d\\
-d&0
\end{array} \right)
\qquad
 \mbox{and}
 \qquad
 T = \left(\begin{array}{cc}
a&c\\
c&b
\end{array} \right) \, \mbox{.}\eqno (7)
$$

Let $x = (x_1, x_2)^{{\rm T}}$ be a standard coordinate column in ${\mathbb{R}}^2$. We introduce the matrices
$$
Z = \left(\begin{array}{cc}
x_1&0\\
x_2&0
\end{array} \right)  \quad \mbox{,} \quad
L = T \cdot Z - \Gamma^{-1} \cdot T \cdot Z \cdot \Gamma  \, \mbox{,}\eqno (8)
$$
$$
B = -\frac{1}{2} T \cdot \Gamma^{-1}  \cdot T = - \frac{\det T}{2} \cdot \Gamma^{-1}  \, \mbox{,}\eqno (9)
$$
where for the last equality we used equality~(6).

Note that from the explicit form~(7) for the matrix $ T $ and equality~(8) by easy calculations we get the following explicit form for the matrix $ L $:
$$
L = \left(\begin{array}{cc}
 ax_1 + cx_2 & cx_1 + bx_2 \\
 cx_1 + bx_2  & -ax_1 -cx_2
\end{array} \right)  \, \mbox{.}
$$
This shows that $L^{{\rm T}} = L$ and ${\mathop {\rm Tr}} L =0$. Thus, the $ L $ matrix is uniquely determined by its first column.

The matrices $ B $ and $ L $ form a Lax pair of dimension $ 2 $ for the system~(1), as the following proposition shows.
\smallskip

{\bf \noindent  {Proposition 3.1.}}  {\it
System ~ (1) is equivalent to the following system
$$
\dot{L} = [B, L]  \, \mbox{.}\eqno (10)
$$
}
\begin{proof}
Notice, that ${\mathop{\rm Tr}} [B,L] =0$ and
$$
[B,L]^{{\rm T}} = (BL -LB)^{{\rm T}}= L^{{\rm T}} B^{{\rm T}} - B^{{\rm T}}L^{{\rm T}}= BL -LB= [B,L]  \, \mbox{,}
$$
therefore, it is sufficient to verify formula~(10) only for the first column of the resulting matrix in the left and right parts of this formula.
Let's calculate
\begin{multline*}
BL -LB = - \frac{1}{2} T \cdot \Gamma^{-1} \cdot T \cdot T \cdot Z +  \frac{1}{2} T \cdot \Gamma^{-1} \cdot T \cdot \Gamma^{-1} \cdot T \cdot Z \cdot \Gamma +  \\ + \frac{1}{2} T \cdot Z \cdot  T \cdot \Gamma^{-1} \cdot T -  \frac{1}{2}\Gamma^{-1} \cdot T \cdot Z \cdot \Gamma
\cdot T \cdot \Gamma^{-1} \cdot T  \, \mbox{.}
\end{multline*}
Since the multiplication of the matrix $ Z $ on the left by any matrix gives a matrix with a zero second column, and the multiplication of the last matrix by the matrix $ \Gamma $
or $\Gamma^{-1}$ on the right gives a matrix with a zero first column, then from the last equality in formula~(9) we find that only the first and fourth term of the sum amount on the first column of  $[B,L]$. The sum of these terms is (taking into account again the second equality in formula~(9)):
\begin{multline*}
 - \frac{1}{2} T \cdot \Gamma^{-1} \cdot T^2 \cdot Z
 -  \frac{\det T}{2}\Gamma^{-1} \cdot T \cdot Z =  - \frac{1}{2} T \cdot \Gamma^{-1} \cdot P \cdot Z
  -  \frac{\det T}{2}  T \cdot T^{-1} \cdot \Gamma^{-1} \cdot T \cdot Z = \\
  =
    - \frac{1}{2} T \cdot \Gamma^{-1} \cdot P \cdot Z   - \frac{1}{2} T \cdot \Gamma^{-1} \cdot T \cdot T \cdot Z
   = -  T \cdot \Gamma^{-1} \cdot P \cdot Z = T \cdot \dot{Z}  \, \mbox{,}
 \end{multline*}
where in the last equality we used the system~(1).
\end{proof}

Let's calculate now  ${\mathop {\rm Tr}}(L^k)$, where $k$ is a natural number.  Note that for any matrix
$L = \left(\begin{array}{cc}
g&h\\
h&-g
\end{array} \right)$
we have $L^2 = (g^2 + h^2) E$, where $E$ is the identity matrix. Therefore,
${\mathop {\rm Tr}}(L^k) =0$, if $k$ is odd, and ${\mathop {\rm Tr}}(L^{2l})= 2(g^2 + h^2)^l$.
In our case, $(g,h)^{{\rm T}} = T x$, where $x = (x_1, x_2)^{{\rm T}}$, $T^2=P$ and $T^{{\rm T}} =T$. Therefore, we obtain
$$
g^2 + h^2 = (T x)^{{\rm T}} \cdot T x =
x^{{\rm T}} T^{{\rm T}} T x = x^{{\rm T}} T^2 x= x^{{\rm T}} Px \, \mbox{.}
$$
So, we get that ${\mathop {\rm Tr}}(L^{2l})= 2(x^{\rm T} P x)^l= 2(2H)^l$, where $H$ is the Hamiltonian of system~(1).

\subsection{Computer calculations}

We now consider the general problem of finding Lax pairs of dimension $ 2 $ for systems of the form
$$
\dot{x}= \tilde{\Gamma} \cdot P \cdot x  \, \mbox{,}\eqno (11)
$$
 where $\tilde{\Gamma}$ is a skew-symmetric, and $P$ is an arbitrary (not necessarily symmetrical) real $2 \times 2$ matrices. Without loss of generality (changing the matrix $P$), we assume that
$$
\tilde{\Gamma} =
\left (\begin{array}{cc}
0&1\\
-1&0
\end{array}
\right ) \qquad \mbox{,} \qquad
P=\left (\begin{array}{cc}
p_1&p_2\\
p_3&p_4
\end{array}
\right ) \, \mbox{.}\eqno (12)
$$

We will look for the matrices $ L $ and $ B $ such that the matrix $ L $ depends linearly on the coordinates $ x $ in the space ${\mathbb{R}}^2$ (or ${\mathbb{C}}^2$), and the matrix $ B $ does not depend on the coordinates $ x $, and the system~(11) implies the system
$$
 \dot{L}=[B,L] \, \mbox{.}\eqno (13)
$$
 Let the first and second columns of the matrix $ L $ be respectively
 $$\left (
\begin{array}{cc}
 {a_1} & {a_2} \\
 {a_3} & {a_4} \\
\end{array} \right) \cdot x
 \qquad \mbox{and} \qquad
 \left(
\begin{array}{cc}
 {y_1} & {y_2} \\
 {y_3} & {y_4} \\
\end{array}
\right) \cdot x  \, \mbox{.} $$
 Let the matrix $ B $ be
$$
B=\left(
\begin{array}{cc}
 {b_1} & {b_2} \\
 {b_3} & {b_4} \\
\end{array}
\right) \, \mbox{.}$$

Then the matrix equation~(13) can be rewritten as a system of $ 8 $ equations in $ 12 $ variables $a_i$, $y_i$, $b_i$ (while the coefficients $ a_i $ and $ y_i $ are determined up to multiplication by a constant). This system is amenable to research using a computer. In particular, it is possible to calculate its Gr\"obner basis. For example, the first element of this basis, which depends only on $ y_3 $ and $ y_4 $, has the form
\begin{multline*}
p_1^2 \, {p_2} \,  {p_4} \, y_4^2- {p_1^2}  \, {p_3} \, {p_4} \, {y_4^2}- {p_1} \,  {p_2^2} \,  {p_3} \,
 {y_4^2}- {p_1} \,  {p_2^2} \,  {p_4} \,  {y_3} \,  {y_4}+ \\ +
{p_1} \, {p_2}  \, {p_3^2} \,  {y_4^2}+ {p_1} \,  {p_2} \,  {p_4^2} \,  {y_3^2}+
{p_1} \,  {p_3^2} \,  {p_4}  \, {y_3} \,  {y_4}- {p_1} \, {p_3} \,  {p_4^2} \,  {y_3^2}+ \\ +
{p_2^3}  \,  {p_3} \, {y_3} \,  {y_4}- {p_2^2}  \, {p_3} \,  {p_4}  \,  {y_3^2}-
{p_2} \,  {p_3^3} \, {y_3} \,  {y_4}+ {p_2} \,  {p_3^2} \, {p_4}  \,  {y_3^2}  \, \mbox{.}
\end{multline*}
(The remaining elements of the Gr\"obner basis, computed using a computer, have a longer record.)

If we additionally assume that the matrix $ P $ is symmetric, that is, $ {p_2 = p_3} $, then the Gr\"obner basis is simplified and the system has the following general solution:
$$
a_1=-y_3 \quad \mbox{,}  \quad a_2=-y_4 \quad \mbox{,} \quad
a_3=\frac{ {p_1} \, {y_4} \, ( {y_1} \, {y_4}-2 \, {y_2} \, {y_3})+2 \, {p_2} \, {y_2} \,  {y_3}^2- {p_4} \, {y_1} \,  {y_3^2}}{
 {y_2} \, ( {p_1} \, {y_2}-2 \, {p_2} \, {y_1})+ {p_4} \, {y_1^2} } \, \mbox{,}\eqno (14)
$$
$$
 a_4=\frac{ {y_4^2} \, (2 \, {p_2} \, {y_1}- {p_1} \, {y_2})+ {p_4} \, {y_3} \, ( {y_2} \, {y_3}-2 \, {y_1} \, {y_4})}{ {y_2} \, ( {p_1}
  \, {y_2}-2 \,  {p_2} \,  {y_1})+ {p_4}  \, {y_1^2}}  \, \mbox{,}\eqno (15)
$$
$$
 {b_2}=-\frac{ {p_1} \, {y_2^2}-2 \, {p_2} \, {y_1} \, {y_2}+ {p_4} \, {y_1^2}}{2 \, ( {y_2} \,  {y_3}- {y_1} \,  {y_4})} \quad \mbox{,}
 \quad {b_3}=\frac{ {p_1} \, {y_4^2}-2 \,  {p_2} \,  {y_3} \,  {y_4}+ {p_4} \, {y_3^2}}{2\,  ( {y_2} \, {y_3}- {y_1} \, {y_4})}
\, \mbox{,}\eqno (16)
$$
$$
 {b_1}=\frac{- {b_4} \, {y_1} \, {y_4}+ {b_4} \, {y_2} \, {y_3}+ {p_1} \,  {y_2} \,  {y_4}-
  {p_2} \,  {y_1} \,  {y_4}- {p_2} \,  {y_2} \,  {y_3}+ {p_4} \,  {y_1} \, {y_3}}{ {y_2} \,  {y_3}- {y_1} \,  {y_4}}  \, \mbox{,}\eqno (17)
$$
dependent on free variables $b_4$, $y_1$, $y_2$, $y_3$, $y_4$. The matrix $ L $ has the form
$$
L=\left(
\begin{array}{cc}
 -x_1  \, {y_3}-x_2 \, {y_4} & \,  x_1 \, {y_1}+x_2 \, {y_2} \\
 q & \, x_1 \, {y_3}+x_2 \, {y_4} \\
\end{array}
\right) \quad \mbox{, where}
$$
\begin{multline*}
q= \frac{x_1 \,(- {p_4} \, {y_1} \, {y_3}^2+2 \, {p_2} \, {y_2} \, {y_3^2}+ {p_1} \, {y_4} \, ( {y_1} \, {y_4}-2 \, {y_2} \,
 {y_3}))}{ {p_4} \, {y_1}^2+ {y_2} \, ( {p_1} \, {y_2}-2 \, {p_2} \, {y_1})}+\\ +
\frac{x_2 \, ((2 \, {p_2} \, {y_1}- {p_1} \,  {y_2}) \, {y_4}^2+ {p_4} \, {y_3} \, ( {y_2} \, {y_3}-2 \, {y_1} \, {y_4}))}{ {p_4} \,
 {y_1}^2+ {y_2} ( {p_1} \,  {y_2}-2 \, {p_2} \,  {y_1})}  \, \mbox{.}
\end{multline*}
The trace of the matrix $ L^2 $ is
$$
{\mathop {\rm Tr}}(L^2)=\frac{4 ( {y_2} \, {y_3}- {y_1} \, {y_4})^2  \,  H }{ {y_2} \, ( {p_1} \, {y_2}-2 \,  {p_2} \, {y_1})+ {p_4} \, {y_1^2}} \, \mbox{,}
$$
where $H = \frac{1}{2} x^{\rm T} P x$ is the Hamiltonian of system~(1).

Note that the denominators in the formulas~(14) - (17) have the following simple interpretation:
$$
{y_2} \, ( {p_1} \, {y_2}-2 \, {p_2} \, {y_1})+ {p_4} \,  {y_1}^2 =(y_1,y_2) \, \tilde{\Gamma}^{\rm T} P \tilde{\Gamma} \, (y_1, y_2)^{\rm T} \, \mbox{,}
$$
$$
y_2 \, y_3-y_1 \, y_4 = (y_1,y_2) \, \tilde{\Gamma}^{\rm T} \, (y_3, y_4)^{\rm T} \, \mbox{.}$$

\smallskip

{\bf \noindent  {Remark 3.2.}}   Note that the Gr\"obner basis is simplified not only for symmetric matrices. Consider the following example.
Let the matrices $ \tilde {\Gamma} $ and $ P $ be as in the formula~(12), moreover $p_1 = p_4 \ne 0$, $p_2 = - p_3 \ne 0$.
For such matrices, the Gr\"obner basis, computed using a computer, is simplified, and all Lax pairs can be found. For example, the matrix $ L $ up to multiplication by a constant has the form
$$
L=\left(
\begin{array}{cc}
 i x_1-x_2 & (x_2-i x_1) y_2 \\
 \frac{i x_1-x_2}{y_2} & x_2-i x_1 \\
\end{array}
\right) ,
$$
where $y_2$ is a free variable.
 However, $ L^2 = 0 $ is always satisfied, therefore these Lax pairs do not give any non-trivial first integrals! (Confer remark~4.4 below, where ${\mathop {\rm Tr}} (L^2) =0$.)

\section{Lax pairs in general case}

In this section, we construct Lax pairs of dimension $ 2 $ for an arbitrary natural number $ n $ for the system~(1).

Suppose that the conditions on the matrices $ \Gamma $ and $ P $ are satisfied, as at the beginning of  Introduction.

Let $\lambda \in {\mathbb{C}}$ be an eigenvalue of the matrix $P \Gamma^{-1}$. Define a subspace $V_{\lambda}  \subset {\mathbb{C}}^{2n}$:
$$
V_{\lambda} = \left\{  x \in {\mathbb{C}}^{2n}  \; : \;  (P \Gamma^{-1})^2 x = \lambda^2 x                         \right\}   \, \mbox{.}
$$
We have that $ \lambda \ne 0 $, and from item~1 of proposition~2.1 we get that $ - \lambda $ is also an eigenvalue of the matrix $ P \Gamma^{- 1}$.
Therefore, $\dim_{{\mathbb{C}}} V_{\lambda}  \ge 2$, since the eigenvectors of the matrix $P \Gamma^{-1}$ with eigenvalues  $ \lambda $ and $ - \lambda $ are linearly independent over ${\mathbb{C}}$ and belong to the subspace  $V_{\lambda}$. Note that if the spectrum of the matrix  $P \Gamma^{-1}$ is simple, then ${\dim_{{\mathbb{C}}} V_{\lambda} =2}$.

  Note that $P \Gamma^{-1} V_{\lambda}  = V_{\lambda}$, since $[P \Gamma^{-1}, (P \Gamma^{-1})^2]=0$ and $\det (P \Gamma^{-1}) \ne 0$.

  Choose an arbitrary non-zero column vector
  $$w = (w_1, \ldots, w_{2n})^{\rm T} \in V_{\lambda} \subset {\mathbb{C}}^{2n} \, \mbox{,}$$ such that  $w$ is not an eigenvector of $P \Gamma^{-1}$.

  We will call such a pair $ (\lambda, w) $ {\em admissible}.

 Define a column vector
  $$
  \hat{w} = i \lambda^{-1} P \Gamma^{-1} w   \; \in \; V_{\lambda} \subset {\mathbb{C}}^{2n}  \, \mbox{,}
  $$
where as usual $i^2=-1$.

Define now
$$
a = x^{\rm T} w = \sum_{1 \le j \le 2n} x_j w_j \quad \mbox{,} \quad  d= x^{\rm T} \hat{w}= i \lambda^{-1} x^{\rm T} P \Gamma^{-1} w = -i \lambda^{-1} (\Gamma^{-1} P x)^{\rm T} w \, \mbox{.}\eqno (18)
$$

We introduce matrices $ B $ and $ L $ of size $ 2 \times 2 $ as follows:
$$
B = -  \frac{i \lambda}{2} \left(
\begin{array}{cc}
 0 & 1 \\
 -1 & 0 \\
\end{array}
\right) \quad \mbox{,} \quad
 L = \left(
\begin{array}{cc}
a & d \\
 d & -a \\
\end{array}
\right) \, \mbox{.} \eqno (19)
$$
\smallskip

{\bf \noindent {Definition 4.1.}} {\em
We define a quadratic function from $\mathbb{C}^{2n}$ to ${\mathbb{C}}$:
$$I_{\lambda, w} = {\mathop{\rm Tr}}(L^2) = 2((x^{\rm T} w)^2 +(x^{\rm T} \hat{w})^2) \, \mbox{,}$$
where the last equality follows from the explicit form~(19) of the matrix $L$.
}

 It is easy to see that (see the end of 3.1) ${\mathop {\rm Tr}} (L^k) =0$, if $k$ is odd, and ${\mathop {\rm Tr}} (L^{2l})= 2(I_{\lambda,w}/2)^l$.

Each so constructed pair $ B, L $
defines a Lax pair of dimension $ 2 $ with interesting properties, as the following theorem shows.

\smallskip

{\bf \noindent {Theorem 4.2.}} {\em
We have properties.
\begin{enumerate}
\item
The pair $ B, L $ is a Lax pair of dimension $ 2 $ for the system~(1).
\item   Let $(\lambda_1, w_1)$ and $(\lambda_2, w_2)$ be admissible pairs, and assume  $\lambda_1^2 \ne \lambda_2^2$. Then
    $$
     (\mathop{\rm grad} I_{\lambda_1, w_1})^{{\rm T}} \cdot \Gamma^{-1} \cdot  \mathop{\rm grad} I_{\lambda_2, w_2} =0  \, \mbox{.}
    $$
 \item \label{t3}   Let $(\lambda_j, w_j)$, where $1 \le j \le l\le n$,  be admissible pairs and all numbers $\lambda_j^2$ are pairwise distinct. Then the complex $1$-forms $d I_{\lambda_j, w_j}$, where $1 \le j \le l$, are linearly independent over  $\mathbb{C}$
 on the complement of the union of at most $ l $ linear subspaces of codimension $ 2 $ in  $\mathbb{C}^{2n}$.
\end{enumerate}
}
\smallskip

{\bf  \noindent {Remark 4.3.}}
Note that if $ (\lambda, w) $ is an admissible pair, then the complex conjugate pair  $(\overline{\lambda}, \overline{w} )$ is also an  admissible pair. Besides,  $\overline{I_{\lambda,w}}= I_{\overline{\lambda}, \overline{w}}$.
Therefore, for real and imaginary parts we have
$$
\mathop{\rm Re} I_{\lambda,w} = \frac{1}{2}(I_{\lambda,w} + I_{\overline{\lambda}, \overline{w}})       \quad \mbox{,}  \quad \mathop{\rm Im}  I_{\lambda,w} = \frac{1}{2i} (I_{\lambda,w} - I_{\overline{\lambda}, \overline{w}})  \, \mbox{.}
$$
Therefore, from section~2 (in particular, from formulas~(5)) and items~2 and 3 of theorem~4.2, taking, if necessary, the real and imaginary parts of the quadratic functions $ I_{\lambda, w} $, we immediately obtain quadratic, functionally independent first integrals for the system~(1), which are pairwise in involution with respect to the Poisson bracket for the Hamiltonian system~(1).
In particular, if the spectrum of the matrix $ P \Gamma^{- 1} $ is simple, then we get $ n $ of such first integrals. This, in particular, gives a new proof of the result from~[5] (as we already noted in  Introduction).

We note here the difference in approaches: to prove the functional independence of the set of first integrals in [5]\footnote{A similar approach using the iteration of various cases is also contained in [6].} a rather cumbersome theorem was used from~[7] on the classification of a pair of bilinear real forms: symmetric and skew-symmetric (the formulation of this theorem is also given in~[8, Appendix~6]). In our case, we construct a set of $ n $ first quadratic integrals in an involution and give a very simple proof (see below), without the classification theorem for a pair of forms,  that they are functionally independent. It follows from the general theory of Lie algebras that all families of first quadratic integrals in an involution generate the same vector space, and, consequently, the first integrals from~[5] are also functionally independent. We will explain the latter in more detail.

Let $ M $ and $ N $ be two real symmetric matrices defining quadratic functions on the space ${\mathbb{R}}^{2n}$.
The Poisson bracket constructed from the symplectic form on ${\mathbb{R}}^{2n}$ defined by the skew-symmetric matrix $ - \Gamma $, being applied to two quadratic functions will again be a quadratic function. This bracket will be rewritten on the set of symmetric matrices as follows:
$$
\{M,N\} \longmapsto -2(M\Gamma^{-1}N-N\Gamma^{-1}M).
$$
Note that the map $M\mapsto -2M\Gamma^{-1}$ defines an isomorphism between the Lie algebra of symmetric matrices with respect to the bracket described above, and the matrix Lie subalgebra consisting of matrices $ B $ satisfying the condition $B^{\rm T}\Gamma^{-1}+\Gamma^{-1}B=0$. The last algebra is isomorphic to the symplectic Lie algebra $sp(2n,{\mathbb{R}} )$ through the map $B\mapsto C^{\rm T}B(C^{\rm T})^{-1}$, where $-C^{\rm T}\Gamma C$ is a standard symplectic unity.

If $ P / 2 $ is a symmetric matrix defining a quadratic Hamiltonian and $ P \Gamma^{- 1} $ has a simple spectrum, then the matrix $-C^{\rm T}P\Gamma^{-1}(C^{\rm T})^{-1}$
also has a simple spectrum, and therefore defines a regular element in the simple Lie algebra $sp(2n,{\mathbb{R}} )$. Then the Cartan subalgebra containing this element consists of all elements commuting with it and is an Abelian subalgebra of dimension $ n $. Therefore, there exists a unique real linear space of dimension  $ n $ consisting of quadratic functions that are in involution with respect to the Poisson bracket and containing the Hamiltonian.

Explicit formulas connecting our first integrals and the first integrals from [2,\,5] are given in [9].

\begin{proof} We prove Theorem~4.2.

1. We note that
$$
\left(
\begin{array}{cc}
 0 & 1 \\
 -1 & 0 \\
\end{array}
\right) \cdot
 \left(
\begin{array}{cc}
a & d \\
 d & -a \\
\end{array}
\right) =
\left(
\begin{array}{cc}
d & -a \\
 -a & -d \\
\end{array}
\right)  \quad \mbox{and}  \quad
\left(
\begin{array}{cc}
 a & d \\
 d & -a \\
\end{array}
\right) \cdot
 \left(
\begin{array}{cc}
0 & 1 \\
 -1 & 0 \\
\end{array}
\right) =
\left(
\begin{array}{cc}
-d & a \\
 a & d \\
\end{array}
\right)
$$
Therefore, the matrices  $BL$, $LB$ and $[B,L]$
will be again symmetric matrices with zero trace. Consequently
we have
$$
[B, L] = B L - L B = B L - (L B)^{\rm T}= 2 B L  \, \mbox{.}
$$
Hence we obtain that system~(4) is equivalent to the following system of equations:
$$
\left\{ \begin{array}{ccc}
\dot{a} & = &  - i \lambda d \\
 \dot{d} & = & i \lambda a \\
\end{array} \right. \eqno (20)
$$
From system~(1) we obtain that
$$
\dot{a}= \dot{x}^{\rm T} \cdot w = (- \Gamma^{-1} P x)^{\rm T} \cdot w  \qquad \mbox{and}  \qquad
\dot{d}= \dot{x}^{\rm T} \cdot \hat{w}= (- \Gamma^{-1} P x)^{\rm T}  \cdot (i \lambda^{-1} P \Gamma^{-1} w)  \, \mbox{.}
$$
Hence we have
$$
\left\{ \begin{array}{ccc}
(- \Gamma^{-1} P x)^{\rm T} \cdot w  & = &  - i \lambda x^{\rm T}  \cdot (i \lambda^{-1} P \Gamma^{-1} w) \\
 (- \Gamma^{-1} P x)^{\rm T} \cdot (i \lambda^{-1} P \Gamma^{-1} w)  & = & i \lambda x^{\rm T} \cdot w  \, \mbox{.}\\
\end{array} \right.
$$
Therefore we obtain
$$
\left\{
\begin{array}{ccc}
x^{\rm T} P \Gamma^{-1} w & = & - i \lambda x^{\rm T}  \cdot (i \lambda^{-1} P \Gamma^{-1} w) \\
x^{\rm T} P \Gamma^{-1} \cdot (i  \lambda^{-1} P \Gamma^{-1} w) & = &  i \lambda x^{\rm T} \cdot w  \, \mbox{.}
\end{array}  \right.
$$
The first equation from the last system is always satisfied. In order for the second equation to be satisfied, it suffices to prove that
$$
(P \Gamma^{-1})^2 w = \lambda^2 w \, \mbox{.}
$$
But we chose the vector  $w \in V_{\lambda}$ with such a condition.

2. From Definition~4.1 we obtain
$$
I_{\lambda, w}=
2 ((x^{\rm T}  w)^2 + (x^{\rm T}  \hat{w})^2)= 2(x^{\rm T}  w + i x^{\rm T}  \hat{w} ) \cdot (x^{\rm T}  w -
i x^{\rm T}  \hat{w}) =
$$
$$
= 2(x^{\rm T}(E - \lambda^{-1} P \Gamma^{-1}  ) w )\cdot (x^{\rm T }(E  + \lambda^{-1} P \Gamma^{-1} )w) \, \mbox{,}\eqno (21)
$$
where  $E$  is the unit matrix.
We note that from this formula we immediately obtain that  ${I _{\lambda, w} \ne 0}$,
since  $w$
is not an eigenvector of the matrix  $P \Gamma^{-1}$.
Now we calculate  $\mathop{\rm grad} I_{\lambda, w}$:
\begin{multline*}
\mathop{\rm grad} I_{\lambda, w} =
2(\mathop{\rm grad}(a^2)  +  \mathop{\rm grad}(d^2) )=
2(2a \cdot \mathop{\rm grad} a    + 2d  \cdot \mathop{\rm grad} d )=
4((x^{\rm T} w) \cdot w + (x^{\rm T} \hat{w}) \cdot \hat{w})=
\\=
4( (x^{\rm T} w) \cdot E - \lambda^{-2} (x^{\rm T} P \Gamma^{-1} w) \cdot P \Gamma^{-1}   )
w =(g_1 \cdot E - g_2 \cdot P \Gamma^{-1}) w  \, \mbox{,}
\end{multline*}
where  $E$  is the unit matrix, and linear functions  $g_1$ and $g_2 $ on ${\mathbb{C}}^{2n}$ are the following:
 $$
 g_1 = 4 x^{\rm T} \cdot w  \qquad \mbox{,}  \qquad g_2 = 4 \lambda^{-2} x^{\rm T} \cdot P \Gamma^{-1} \cdot w  \, \mbox{.}
 $$
 Now, since for any  $x \in {\mathbb{C}}^{2n}$ the matrix  $g_1 \cdot E - g_2 \cdot P \Gamma^{-1}$
 commutes with the matrix $(P \Gamma^{-1})^2$,
 we obtain that for any $x \in {\mathbb{C}}^{2n}$
 the vector $\mathop{\rm grad} I_{\lambda, w}$
 is an eigenvector of the matrix  $(P \Gamma^{-1})^2$
 with  eigenvalue $\lambda^2$. Now item~2 of Theorem~4.2 follows from item~2 of Proposition~2.1.

 3. Since the non-zero eigenvectors corresponding to different eigenvalues are linearly independent, the statement of item~3 of this theorem
 follows from the explicit calculation of the gradient
 $\mathop{\rm grad} I_{\lambda, w}$ performed
 during the proof of the previous item. Besides, since the vector $ w $ is not an eigenvector  of the matrix $P \Gamma^{-1}$, the gradient
$\mathop{\rm grad} I_{\lambda, w}$
will be equal to zero only on   the intersection of two linear complex hyperplanes in~${\mathbb{C}}^{2n}$: $g_1 =0$ and $g_2=0$.
\end{proof}

\smallskip

{\bf \noindent Remark 4.4.}
We note if we, when constructing a Lax pair,  would take an eigenvector  $w$ of the matrix $(P \Gamma^{-1})^2$
 such that this eigenvector would be an eigenvector for the matrix  $P \Gamma^{-1}$,
 then we would obtain that  ${\mathop {\rm Tr}}(L^2) =0$ (this follows at once from formula~(21)).
 Hence we obtain the importance of the property  $\dim_{{\mathbb{C}}} V_{\lambda} \ge 2$.
  It is the absence of such a property that prevents us to construct a Lax pair with non-zero function  ${\mathop {\rm Tr}} (L^2)$
  for arbitrary linear system
$$
\dot{x}= C x   \, \mbox{.}
$$
\smallskip

{\bf \noindent Remark 4.5.}
If the spectrum of matrix  $P\Gamma^{-1}$  is simple,
we can construct a Lax pair of dimension  $2n$
 such that system~(4)
 will be equivalent to the system~(1).
 To do this it is enough to take block diagonal matrices  $B$ and $L$
 with blocks that are  matrices of size $2 \times  2$, as in formula~(19),
 with eigenvalues  $\lambda_i$ such that  $\lambda_i^2\neq \lambda_j^2$ when  $i\neq j$.

Indeed, as we see from the proof of item~1 of Theorem~4.2,
system~(4)
is equivalent to systems  (20)
with different pairs  $(\lambda ,w)$. But then the equations of these systems can be rewritten as systems
$$
(\dot{x}+\Gamma^{-1}Px)^T\cdot w=0, \mbox{\quad} (\dot{x}+\Gamma^{-1}Px)^T\cdot \hat{w}=0
$$
for  $n$ different eigenvectors  $w$. Since all such vectors  $w, \hat{w}$ are linearly independent,
we obtain system~(1):
$$
\dot{x}+\Gamma^{-1}Px=0.
$$
\smallskip

{\bf \noindent Remark 4.6.}
We can construct Lax pairs of dimension  $2n$
 in other ways. Let, for example, the matrices  $B$ and $L$
 be from blocks which are  matrices of size $n \times n$.
 We fix an eigenvalue $\lambda$ of the matrix  $P \Gamma^{-1}$.
 We consider the following   $B$ and $L$:
 $$
 B=- \frac{i\lambda}{2} \left(
\begin{array}{cc}
 0 & E \\
 -E & 0 \\
\end{array} \right) \qquad \mbox{,}
 \qquad
 L =\left(
\begin{array}{cc}
 A & D \\
 D & -A \\
\end{array} \right)  \, \mbox{,}$$
 where $E$ is the unit matrix of size  $n \times n$, and  $A$ and $D$ are symmetric matrices of size  $n \times n$.
  Let $A = (a_{kl})_{1 \le k,l\le n}$, $D = (d_{kl})_{1 \le k,l \le n}$.
  Then it is enough to specify the matrix elements $a_{k,l}$ and $d_{k,l}$ when $l \ge k$.
  For every such pair  $k,l$
  we define  $a_{k,l}$ and $d_{k,l}$ in the same way  as
   we defined $a$ and $d$ (see formula~(18)) when we constructed  the Lax pair.
    Moreover,  the eigenvalue $\lambda$
    must be the same, and the eigenvectors  $w \in V_{\lambda}$ can change.
    For such a Lax pair the function  ${\mathop {\rm Tr}}(L^2)$
    is the sum of the functions  $I_{\lambda, w}$, which  we studied above.

Moreover, we can write the Lax pair of dimension  $2n$
with real coefficients and such that several eigenvalues of the matrix
$P \Gamma^{-1}$ are taken into account.
We consider the matrix  $B$ of the form
$
 B=- \frac{1}{2} \left(
\begin{array}{cc}
 0 & J \\
 -J & 0 \\
\end{array} \right) \, \mbox{,}
$
where
$J$
is a diagonal matrix with entries  $i\lambda_1, \ldots , i \lambda_n $
on the diagonal,
and for any  $1 \le j \le n $  the number  $\lambda_j$
is the eigenvalue of the matrix $P \Gamma^{-1}$
 such that if $\lambda_j^2 \notin {\mathbb{R}} $, then   $\overline{\lambda}_j = \lambda_k$
 for some  $k \ne j $.
 Now we take the matrix  $L$ as above, but with diagonal matrices  $A$ and $D$
 such that the $j$-th numbers on diagonals of matrices  $A$ and $D$
 are constructed  as numbers  $a$  and $d$ in formula~(18) for the eigenvalue  $\lambda _j$.
 Then the pair  $B$,
  $L$
  gives a Lax pair of dimension $2n$  for system~(1), and
   the function   ${\mathop {\rm Tr}}(L^2)$
   is the sum of the functions
    $I_{\lambda_j, w}$ introduced above.
    By conjugating matrices $B$ and $L$  by means of permutation matrices (that is, by rearranging the vectors in the basis by some permutation),
    we obtain a block diagonal matrix  $B$ with blocks
$-\frac{1}{2}\left(
\begin{array}{cc}
 0 & i\lambda \\
 -i\lambda & 0 \\
\end{array} \right)
$
of size $2 \times 2$.
If  $\lambda^2 \in {\mathbb{R}} $,
then either the block itself, or its Jordan normal form will be real matrices.
If $\lambda^2 \notin {\mathbb{R}}$,
 then two such blocks with complex conjugate numbers  $\lambda = a \pm bi$
 have all different eigenvalues, and these eigenvalues are the same  as the eigenvalues of the real matrix of size  $4 \times 4$
 which can be written as a block matrix:
$$\left(
\begin{array}{cc}
 0 & N_1 \\
 N_2 & 0 \\
\end{array} \right) \, \mbox{, where} \quad
 N_1 =
 \left(
\begin{array}{cc}
 a & b \\
 b & -a \\
\end{array} \right)
\quad \mbox{and} \quad
N_2 =
 \left(
\begin{array}{cc}
 a & -b \\
 -b & -a \\
\end{array} \right) \, \mbox{.}
$$
Therefore we can make a suitable replacement of the basis, and the matrix  $B$ will be real.
Then the resulting Lax pair can be divided into two real ones: with the real and imaginary parts of the matrix  $L$.
Since the trace does not change under the conjugation of a matrix, we obtain the same integrals  as for the original Lax pair  (the real and imaginary part  of the trace of  $L^2$).
\smallskip

{\bf \noindent Example 4.7.}
In the case $n=1$,  we can always ``normalize'' (that is, multiple by a complex number) the vector $w$
from an admissible pair $(\lambda,w)$  in such a way that  $I_{\lambda, w}=4H$,
 where  $H$  is the Hamiltonian of system~(1).
 Namely, the vector  $w$  must satisfy a condition:
$$
w^{\rm T}\Gamma P\Gamma^{-1} w=-\lambda^2 \cdot \det \Gamma= \det P  = p_1p_4-p_2^2  \, \mbox{,} \eqno (22)
$$
where the coefficients in the matrix  $P$
are denoted as in the right part of formula~(12) with the additional condition  $p_3=p_2$.
(Since the vector $P \Gamma^{-1} w$ is not proportional to the vector  $w$, the left side in formula~(22)  is not equal to zero.)
In this case, by simple direct calculations we obtain:
\begin{gather*}
a^2=(w_1x_1+w_2x_2)^2 \quad \mbox{,}  \quad
d^2=\frac{(w_2 ({p_1} {x_1}+{p_2}{x_2})+{w_1} (-{p_2} {x_1}-{p_4} {x_2}))^2}{(p_1p_4-p_2^2)}  \, \mbox{,}  \\
I_{\lambda, w}=2(a^2 + d^2)=2(x_1^2(p_1 p_4 w_1^2-2 p_1 p_2 w_1 w_2+p_1^2 w_2^2)+ \\ +
2x_1x_2( p_2 p_4 w_1^2-2 p_2^2 w_1 w_2+ p_1 p_2 w_2^2) + x_2^2(p_4^2 w_1^2-2 p_2 p_4 w_1 w_2+p_1 p_4 w_2^2))/(p_1p_4-p_2^2)= 4H
\end{gather*}

We note also that the condition in formula~(22)
can be rewritten as
  ${w^{\rm T} P^{-1} w =1}$ by means of identity~(6) (where $T$ must be replaced by  $P$).

Besides, in the notation of section~3.2, the corresponding Lax pair from Theorem~4.2 is obtained if we take  $b_4=0$, $(y_3,y_4)^T=-w$, $(y_1,y_2)^T=\hat{w}$.
\smallskip

{\bf \noindent Example 4.8.}
In the case $n=2$ we consider, for example, the following matrices:
$$
P=\left(
\begin{array}{cccc}
 0 & a & 0 & b \\
 a & 0 & -b & 0 \\
 0 & -b & 0 & a \\
 b & 0 & a & 0 \\
\end{array}
\right), \mbox{\quad}
\Gamma =\left(
\begin{array}{cccc}
 0 & 1 & 0 & 0 \\
 -1 & 0 & 0 & 0 \\
 0 & 0 & 0 & 1 \\
 0 & 0 & -1 & 0 \\
\end{array}
\right)
$$
We define, as in the first part of Remark~4.6,
$$
B=
\left(
\begin{array}{cccc}
 0 & 0 & h & 0 \\
 0 & 0 & 0 & h \\
 -h & 0 & 0 & 0 \\
 0 & -h & 0 & 0 \\
\end{array}
\right) \quad \mbox{,} \qquad
L =
\left(
\begin{array}{cccc}
 x^{\rm T} w & 0 & x^{\rm T} \hat{w} & 0 \\
 0 &  -x^{\rm T} \hat{w} & 0 & x^{\rm T} w \\
 x^{\rm T} \hat{w}  & 0 & -x^{\rm T} w & 0 \\
 0 & x^{\rm T} w & 0 & x^{\rm T}  \hat{w}\\
\end{array}
\right) \, \mbox{,}
$$
where $ h=-i\lambda/2=-\frac{1}{2} (-b+ai)$,   $\lambda = a+ bi $,
and two eigenvectors with the eigenvalue $\lambda^2$ for the matrix  $(P\Gamma^{-1})^2$ are chosen: the firs one is $w=(1,1,i,-i)^{\rm T}$,
the second one is   $\tilde{w}= -\hat{w}=(-i,i,1,1)^{\rm T}$.
We have $\hat{w}= (i, -i,-1,-1)^{\rm T}$,
$\hat{\tilde{w}}=w$.
Then the matrices  $B, L$
form a Lax pair for system~(1).
We define
$$
H_1=x_1x_2+ x_3x_4 \quad \mbox{,} \quad H_2=x_1 x_4 - x_2 x_3 \, \mbox{.}
$$
Then it is not difficult to calculate that
$$
 16 H_1 =
 I_{\lambda,w}+I_{\overline{\lambda},\overline{w}} \quad \mbox{,} \quad
 16 H_2=
 {i}(I_{\lambda,w}-I_{\overline{\lambda},\overline{w}}) \, \mbox{.} \eqno (23)
$$
Hence we obtain that
$$
{\mathop {\rm Tr}}(L^2)
=
4 ( (x^{\rm T}w)^2 + (x^{\rm T} \hat{w})^2  ) = 2I_{\lambda,w} = 16 H_1 - 16 H_2 i  \, \mbox{.}
$$

\section{The maps which appear from constructions of Lax pairs are Poisson}

In this section, we show that on the image of the map arising in Lax pairs from section~4 there is a symplectic structure such that the matrix equation of the Lax pair is Hamiltonian, and the map itself which appear  from the construction of the Lax pair  is Poisson, that is, it preserves the Poisson brackets, where on ${\mathbb{R}}^{2n}$
there is the Poisson bracket induced by the Hamiltonian system~(1), see~section~2.

By an admissible pair $(\lambda, w)$ such that ${w \in \mathbb{R}^{2n}}$ if ${\lambda^2 \in \mathbb{R}}$,
and by the Lax pair from Theorem~4.2 we define a map
$$
\Phi_{\lambda, w} \; : \;   \mathbb{R}^{2n}  \longrightarrow \mathbb{C} e_1 \oplus \mathbb{C} e_3    \quad \mbox{,}  \quad
x \mapsto a(x) \cdot e_1 + d(x) \cdot e_3  \mbox{.}
$$
Depending on three cases : 1)  $\lambda \in i\mathbb{R}$; 2) $\lambda \in \mathbb{R}$; and 3) $\lambda \in \mathbb{C}$, $\lambda^2\notin \mathbb{R}$,
this map defines maps of real spaces for which we keep the same notation:
$$
 \Phi_{\lambda, w} \; : \;   \mathbb{R}^{2n}  \longrightarrow \mathbb{R}  e_1 \oplus \mathbb{R} e_3   \, \mbox{,}
$$
$$
 \Phi_{\lambda, w} \; : \;   \mathbb{R}^{2n}  \longrightarrow \mathbb{R} e_1 \oplus \mathbb{R} ie_3   \quad \mbox{,}  \quad x \mapsto a(x) \cdot e_1 + (i^{-1}d(x)) \cdot i e_3  \, \mbox{,}
$$
\begin{gather*}
\Phi_{\lambda, w} \, : \, \mathbb{R}^{2n}  \to  \mathbb{R} e_1 \oplus \mathbb{R} e_2 \oplus \mathbb{R} e_3 \oplus \mathbb{R} e_4    \;  \mbox{,}  \\
 x \mapsto (x^{{\rm T}}w_1, x^{{\rm T}}w_2, x^{{\rm T}} \hat{w}_1, x^{{\rm T}} \hat{w}_2)^{{\rm T}}  \mbox{,}
\end{gather*}
where $w= w_1 + i w_2$, $ \hat{w}= \hat{w}_1  + i \hat{w}_2 $, $e_2 = i e_1$ and  $e_4 = i e_3$.
\smallskip

{\bf \noindent Theorem 5.1} {\em
Let the spectrum of matrix  $P \Gamma^{-1}$ is simple.
We choose any admissible pair  $(\lambda, w)$ such that $w \in \mathbb{R}^{2n}$ if $\lambda^2 \in \mathbb{R}$.
Then on the image of the map $\Phi_{\lambda, w}$
there is a natural symplectic structure such that the system obtained from the Lax pair is Hamiltonian with respect to this structure, and the map
 $\Phi_{\lambda, w}$ is Poisson.
}
\begin{proof}
We define
$$
K_{\lambda, w} = - w^{\rm T} \Gamma^{-1} \hat{w} = - i \lambda^{-1} w^{\rm T} \Gamma^{-1} P \Gamma^{-1} w = - i \lambda w^{\rm T} P^{-1} w  \,  \mbox{.}
$$
From the conditions of the theorem and item~3  of Proposition~2.1 it follows that  $K_{\lambda,w} \ne 0$.

{\em In case 1)} let $z_1$ and $z_3$
be the real coordinates on vectors  $e_1$ and $e_3$  correspondingly.
To define the Poisson bracket  $\{ \cdot, \cdot \}$
on smooth functions on the space  ${\mathbb{R}}\langle e_1, e_3  \rangle$,
so that the map
$\Phi_{\lambda, w}$ would be a morphism of Poisson manifolds, we define
$$
\{z_1, z_1  \} = \{z_3, z_3  \} =0 \, \mbox{, } \eqno (24)
$$
$$
 \{z_1,z_3\}= -\{z_3, z_1\}= \{\Phi_{\lambda, w}^*(z_1), \Phi_{\lambda, w}^*(z_3)\} =
\{ x^{\rm T} w, x^{\rm T} \hat{w}  \} = - w \Gamma^{-1} \hat{w}  = K_{\lambda,w}  \mbox{,} \eqno (25)
$$
where for functions on ${\mathbb{R}}^{2n}$
the Poisson bracket is given by the skew-symmetric matrix $- \Gamma$, as for the Hamiltonian system~(1).
Now the non-degenerate skew-symmetric matrix
$$
\Gamma_{\lambda, w} = \left(
\begin{array}{cc}
 0& - K_{\lambda, w}^{-1}\\
 K_{\lambda,w}^{-1} & 0 \\
\end{array} \right)
$$
defines the symplectic structure on the space  ${\mathbb{R}} \langle e_1, e_3\rangle$
such that the corresponding Poisson brackets  (see~section~2) will be calculated by formulas~(24)-(25)
which imply that
 for smooth functions  $h_1$ and $h_2$ on ${\mathbb{R}} \langle e_1 , e_3 \rangle$ we have:
$$
\{ h_1, h_2\} = K_{\lambda, w} \frac{\partial h_1}{\partial z_1} \frac{\partial h_2}{\partial z_3} -
K_{\lambda, w} \frac{\partial h_1}{\partial z_3} \frac{\partial h_2}{ \partial z_1}  \, \mbox{.}
$$
And we obtain
$$
\{h_1, h_2  \} = \{ \Phi_{\lambda, w}^*(h_1)   , \Phi_{\lambda, w}^*(h_2)\} \, \mbox{.}
$$

We recall that the matrix equation defined by Lax pair~(4) is equivalent to system~(20), that is, to the system
$$
\left\{ \begin{array}{ccc}
\dot{z_1} & = &  - i \lambda z_3 \\
 \dot{z_3} & = & i \lambda z_1 \\
\end{array} \right.   \, \mbox{,}
$$
which is, as it is easy to see, is equivalent to the system
$$
\Gamma_{\lambda, w} (\dot{z}_1, \dot{z}_3)^{\rm T} =  P_{\lambda, w} (z_1, z_3)^{\rm T} \, \mbox{,}
$$
where $P_{\lambda, w}$
is the scalar matrix with real numbers  $i \lambda K_{\lambda, w}^{-1}$ on the diagonal.
From section~2 we obtain at once that this system is Hamiltonian with respect to the symplectic form given by the skew-symmetric matrix $\Gamma_{\lambda, w}$.

{\em In case  2)}
let   $z_1$ and $z_4$
be the real coordinates on the vectors  $e_1$ and $ie_3$ correspondingly.
We define
$$\{z_1, z_1 \} = \{z_4, z_4  \} =0 \, \mbox{,}$$
$$
\{z_1, z_4  \}= \{  \Phi_{\lambda, w}^* (z_1),  \Phi_{\lambda, w}^*(z_4) \} = \{ x^{\rm T} w, i^{-1}x^{\rm T} \hat{w} \} = i w^{\rm T} \Gamma^{-1} \hat{w}=
-i K_{\lambda , w}  \, \mbox{.}
$$
Now a non-degenerate skew-symmetric matrix
$$
\tilde{\Gamma}_{\lambda,w} =
\left(
\begin{array}{cc}
 0& -i K_{\lambda, w}^{-1}\\
i K_{\lambda,w}^{-1} & 0 \\
\end{array} \right)
$$
gives the symplectic structure on the space   ${\mathbb{R}}\langle e_1, ie_3  \rangle$, so that
for any two smooth functions
 $h_1$ and $h_2$ on the space   ${\mathbb{R}}\langle e_1, ie_3  \rangle$ we have
 $$
\{h_1, h_2\} = \{ \Phi_{\lambda, w}^*(h_1), \Phi_{\lambda, w}^* (h_2)  \}  \, \mbox{.}
$$
Now the matrix equation defined by the Lax pair is equivalent to  system~(20), and in our case this system can be rewritten as
$$
\left\{ \begin{array}{ccc}
\dot{z_1} & = &  \lambda z_4 \\
 \dot{z_4} & = &   \lambda z_1 \\
\end{array} \right.   \,
$$
that is equivalent to the system
$$
\tilde{\Gamma}_{\lambda, w} (\dot{z}_1, \dot{z}_4)^{\rm T} =  \tilde{P}_{\lambda, w} (z_1, z_4)^{\rm T} \, \mbox{,}
$$
where the matrix  $\tilde{P}_{\lambda, w} $
is a diagonal matrix with real numbers
 $-i\lambda K_{\lambda, w}^{-1} $, $i \lambda K_{\lambda, w}^{-1} $
 along the diagonal.
 The last system is Hamiltonian with respect to the symplectic form given by the skew-symmetric matrix
 $\tilde{\Gamma}_{\lambda,w}$.

{\em In case 3)} let $z_k$  be the real coordinate on the vector  $e_k$,
where  $1 \le k \le 4$. Then we have
$$
\Phi_{\lambda, w}^* (z_1)=   x^{\rm T}w_1 \quad \mbox{,}  \quad
\Phi_{\lambda, w}^*(z_2) =  x^{\rm T} w_2  \quad  \mbox{,} \quad
\Phi_{\lambda, w}^*(z_3)=   x^{\rm T} \hat{w}_1  \quad   \mbox{,}  \quad
\Phi_{\lambda, w}^*(z_4) =   x^{\rm T}  \hat{w}_2     \, \mbox{.}
$$
We define the Poisson brackets:
\begin{multline*}
\{ z_1, z_2 \} = - \{z_2, z_1  \}= w_1^{\rm T} (- \Gamma^{-1}) w_2 = \\ =
 (-\frac{w + \overline{w}}{2})^{\rm T} \cdot \Gamma^{-1} \cdot \frac{w -\overline{w}}{2i} =
\frac{i}{4}(w^{\rm T} \Gamma^{-1} w - \overline{w}^{\rm T} \Gamma^{-1} \overline{w} -2 w^{\rm T} \Gamma^{-1} \overline{w}) =0 \,  \mbox{,}
\end{multline*}
where we used that $w^{\rm T} \Gamma^{-1} \overline{w} =0 $ by item~2 of Proposition~2.1,
because $\overline{w}$ is an eigenvector with the eigenvalue $\overline{\lambda}^2$ of the matrix  $(P \Gamma^{-1})^2$, and $\overline{\lambda}^2 \ne \lambda^2$.

By similar calculations we obtain  that
 $\{z_3 , z_4\} = -\{z_4, z_3 \} =0$,
 since  $\hat{w}$ and  $\overline{\hat{w}}$ are eigenvectors of the matrix
$(P \Gamma^{-1})^2$ with eigenvalues  $\lambda^2$ and $\overline{\lambda}^2$ correspondingly.
We calculate now
\begin{multline*}
\{z_1, z_3  \} =
- w_1 \Gamma^{-1} \hat{w}_1
=
-(\frac{w + \overline{w}}{2})^{\rm T} \cdot \Gamma^{-1} \cdot \frac{\hat{w} + \overline{\hat{w}}}{2} = \\ =
-\frac{1}{4} (w^{\rm T} \Gamma^{-1} \hat{w} + \overline{w^{\rm T} \Gamma^{-1} \hat{w}})= \frac{1}{4}(K_{\lambda, w} + \overline{K}_{\lambda, w}) \, \mbox{,}
\end{multline*}
where we used that $\overline{w}^{\rm T} \Gamma^{-1} \hat{w} =0 =  w^{\rm T} \Gamma^{-1} \overline{\hat{w}} $
by item~2 of Proposition~2.1.
Similarly, we calculate
\begin{multline*}
\{z_2, z_4  \} = - w_2 \Gamma^{-1} \hat{w}_2 = -(\frac{w - \overline{w}}{2i})^{\rm T} \cdot \Gamma^{-1} \cdot \frac{\hat{w} - \overline{\hat{w}}}{2i} = \\
=\frac{1}{4} (w^{\rm T} \Gamma^{-1} \hat{w} + \overline{w^{\rm T} \Gamma^{-1} \hat{w}})= - \frac{1}{4}(K_{\lambda, w} + \overline{K}_{\lambda, w}) =
-\{z_1, z_3\} \mbox{.}
\end{multline*}
\begin{multline*}
\{z_1 , z_4 \} =-w_1^{\rm T} \Gamma^{-1}  \hat{w}_2=
 -(\frac{w + \overline{w}}{2})^{\rm T} \cdot \Gamma^{-1} \cdot \frac{\hat{w} - \overline{\hat{w}}}{2i} = \\
=\frac{i}{4} (w^{\rm T} \Gamma^{-1} \hat{w} - \overline{w^{\rm T} \Gamma^{-1} {\hat{w}}})=  \frac{i}{4}(-K_{\lambda, w} + \overline{K}_{\lambda, w}) \, \mbox{.}
\end{multline*}
\begin{multline*}
\{z_2 , z_3 \} =-w_2^{\rm T} \Gamma^{-1}  \hat{w}_1=
 -(\frac{w - \overline{w}}{2i})^{\rm T} \cdot \Gamma^{-1} \cdot \frac{\hat{w} + \overline{\hat{w}}}{2} = \\
=\frac{i}{4} (w^{\rm T} \Gamma^{-1} \hat{w} - \overline{w^{\rm T} \Gamma^{-1} {\hat{w}}})=  \frac{i}{4}(-K_{\lambda, w} + \overline{K}_{\lambda, w}) \, \mbox{.}
\end{multline*}

Thus, the matrix of the pairwise Poisson brackets between the coordinate functions
$z_1, \ldots, z_4$ is the following  $4 \times 4$ real skew-symmetric matrix,
which we write as  a block matrix with blocks of size
 $2 \times 2$:
$$
Y = \left(
\begin{array}{cc}
 0& R \\
- R & 0 \\
\end{array} \right)  \, \mbox{,}
$$
where  $R$ is a real symmetric matrix of size $2 \times 2$ with zero trace. And we have
$$ \hspace{-0.1cm} \det R =  \frac{1}{16}(-(K_{\lambda,w} + \overline{K}_{\lambda, w})^2 + (- K_{\lambda,w} + \overline{K}_{\lambda, w})^2    ) =
\frac{1}{16}((-2K_{\lambda,w}) (2 \overline{K}_{\lambda, w}))= -\frac{1}{4} \mid K_{\lambda, w} \mid^2 \ne 0
  \mbox{.}$$
  We note that   $R^{-1}$
  is again a symmetric with zero trace matrix.
  Now, from the calculations performed, it follows that the skew-symmetric non-degenerate matrix
  $$
  Y^{-1} = \left(
\begin{array}{cc}
 0& -R^{-1} \\
 R^{-1} & 0 \\
\end{array} \right)
  $$
  defines the symplectic structure on the space  ${\mathbb{R}}^4$,
  so that the map    $\Phi_{\lambda, w}$
  will be Poisson with respect to this symplectic structure and the symplectic structure for the system~(1) (see section~2),
  that is, the inverse image map  $\Phi_{\lambda, w}^*$   preserves the Poisson bracket on functions.

Let $\lambda = \alpha_1 + i \alpha_2 $, where $\alpha_1$ and $\alpha_2$ are real numbers.
Then the matrix equation given by  Lax pair~(4)
is equivalent to system~(20), that is, to the system
$$
\left\{ \begin{array}{ccc}
\dot{z}_1 + i \dot{z}_2 & = & (\alpha_2 - i \alpha_1)(z_3 + iz_4)   \\
 \dot{z}_3 + i\dot{z}_4 & = & (-\alpha_2 + i \alpha_1) (z_1 + i  z_2) \\
\end{array} \right.   \, \mbox{,}
$$
The last system is equivalent to the following system:
$$
(\dot{z}_1, \dot{z}_2, \dot{z}_3, \dot{z}_4)^{\rm T} = C (z_1, z_2, z_3, z_4)^{\rm T} \, \mbox{,}\eqno (26)
$$
where the block matrix  $C$ of size  $4 \times 4$ is:
$$
C  = \left(
\begin{array}{cc}
 0& F \\
 -F & 0 \\
\end{array} \right)
\quad \mbox{, where $2 \times 2$-matrix $F$ is:} \quad
F = \left(
\begin{array}{cc}
 \alpha_2& \alpha_1 \\
 -\alpha_1 & \alpha_2 \\
\end{array} \right)  \, \mbox{.}
$$
Since the product of a symmetric matrix  of size
$2 \times 2$ with zero trace on matrix  $F$
is again a symmetric matrix with zero trace, the matrix ${Q = Y^{-1} C}$
is a symmetric block diagonal matrix of the form
$$
Q =\left(
\begin{array}{cc}
 M& 0 \\
 0 & M \\
 \end{array}
 \right)
 \,  \mbox{,}
$$
where $M$
is a real symmetric matrix of size $2 \times 2$ with zero trace, and  $\det M \ne 0$.
Therefore system~(26)  is equivalent to system
$$
-Y^{-1} (\dot{z}_1, \dot{z}_2, \dot{z}_3, \dot{z}_4)^{\rm T}= - Q (z_1, z_2, z_3, z_4)^{\rm T}  \mbox{,} \eqno (27)
$$
which,  according to section~2, is Hamiltonian with the symplectic structure defined by the skew-symmetric matrix
 $Y^{-1}$.
\end{proof}

From theorem~5.1 and item~2 of Proposition~2.1  we obtain a corollary.
\smallskip

{\bf \noindent Corollary 5.2. } {\em
Under conditions of Theorem 5.1, we consider several admissible pairs  $(\lambda_j, w_j)$, where $1 \le j \le l$,
such that  $\lambda_{j_1}^2 \ne \lambda_{j_2}^2 $ and $\lambda_{j_1}^2 \ne \overline{\lambda_{j_2}^2}$
for any natural numbers  $1 \le j_1 < j_2 \le l$.
We define on the image of the map $\prod_{j=1}^l \Phi_{\lambda_j, w_j}$
 the block diagonal symplectic structure, where each block corresponds to the symplectic structure for the pair  $(\lambda_j,w_j)$
from Theorem~5.1.  Then the map  $\prod_{j=1}^l \Phi_{\lambda_j, w_j}$ is Poisson.
}

\vspace{0.5cm}

\noindent
D. V. Osipov

{\small
\noindent Steklov Mathematical Institute of Russsian Academy of Sciences, 8 Gubkina St., Moscow 119991, Russia, {\em and}

\noindent National Research University Higher School of Economics, Laboratory of Mirror Symmetry, NRU HSE, 6 Usacheva str., Moscow 119048, Russia,
{\em and}

\noindent National University of Science and Technology ``MISiS'',  Leninsky Prospekt 4, Moscow  119049, Russia
}

\noindent {\it E-mail:}  ${d}_{-} osipov@mi{-}ras.ru$

\vspace{0.3cm}

\noindent
A. B. Zheglov

\noindent  Lomonosov Moscow State University, faculty of mechanics and mathematics, department of differential geometry and applications, Leninskie gory, GSP, Moscow, 119899, Russia

\noindent {\it E-mail:}  $azheglov@math.msu.su$

\end{document}